\documentclass{aa}  

\usepackage{graphicx}
\usepackage{txfonts}
\begin{document} 

   \title{Searching for Anomalous Microwave Emission in nearby galaxies}
   \subtitle{K-band observations with the Sardinia Radio Telescope}

   \author{
   S.~Bianchi\inst{1}
   \and
   M.~Murgia\inst{2}
   \and
   A.~Melis\inst{2}
   \and
   V.~Casasola\inst{3}
   \and
   F.~Galliano\inst{4}
   \and
   F.~Govoni\inst{2}
   \and 
   A.~P.~Jones\inst{5}
   \and 
   S.~C.~Madden\inst{4}
   \and \\
   R.~Paladino\inst{3}
   \and 
   F.~Salvestrini\inst{1}
   \and 
   E.~M.~Xilouris\inst{6}
   \and
   N.~Ysard\inst{5}
}

\institute{
INAF-Osservatorio Astrofisico di Arcetri, L. E. Fermi 50, 50125, Firenze, Italy\\
\email{simone.bianchi@inaf.it}
\and
INAF-Osservatorio Astronomico di Cagliari, Via della Scienza 5, 09047 Selargius (CA), Italy
\and
INAF-Istituto di Radioastronomia-Via P. Gobetti, 101, 40129 Bologna, Italy
\and 
AIM, CEA, CNRS, Universit\'e Paris-Saclay, Universit\'e  Paris Diderot, Sorbonne Paris Cit\'e, 91191 Gif-sur-Yvette, France
\and
Universit\'e  Paris-Saclay, CNRS, Institut d'Astrophysique Spatiale, 91405 Orsay, France
\and
National Observatory of Athens, Institute for Astronomy, Astrophysics, Space Applications and Remote Sensing, Ioannou Metaxa and Vasileos Pavlou, 15236 Athens, Greece
}

\date{}

 
\abstract
{}
{We observed four nearby spiral galaxies (NGC~3627, NGC~4254, NGC~4736 and NGC~5055) in the K band with the 64-m Sardinia Radio Telescope, with the aim of detecting the Anomalous Microwave Emission (AME), a radiation component presumably due to spinning dust grains, observed so far in the Milky Way and in a handful of other galaxies only (most notably, M~31).}
{We mapped the galaxies at 18.6 and 24.6~GHz and studied their global photometry together with other radio-continuum data from the literature, in order to find AME as emission in excess of the synchrotron and thermal components.}
{
We only find upper limits for AME. 
These non-detections, and other upper limits in the literature, are nevertheless consistent with 
the average AME emissivity from the few detections: it is
$\epsilon^\mathrm{AME}_{\mathrm{30~GHz}} = 2.4\pm0.4 \times 10^{-2}$ MJy sr$^{-1}$ (M$_\odot$ pc$^{-2}$)$^{-1}$
in units of dust surface density (equivalently, 
$1.4\pm0.2 \times 10^{-18}$ Jy sr$^{-1}$ (H cm$^{-2}$)$^{-1}$
in units of H column density).
We finally suggest to search for AME in quiescent spirals with relatively low radio luminosity, such as M~31.}
{}

 
\titlerunning{}
   \maketitle
%

\section{Introduction}

Observations in the Milky Way (MW) at $10 \la \nu/\mathrm{GHz} \la 50$ have revealed emission 
in excess of the well known thermal dust, free-free and synchrotron radiation components: it is the 
Anomalous Microwave Emission (AME), peaking at $\nu\approx30$~GHz and correlating with thermal dust 
emission at higher frequencies (for a review, see \citealt{DickinsonNAR2018}). AME is seen both in the 
diffuse medium (\citealt[][hereafter PLA16]{Planck2015XXV}) as well as on isolated clouds, where 
its characteristics are reported to vary with environment \citep{PlanckIntermediateXV}.

The most popular explanation for AME involves electric dipole emission from spinning dust grains \citep{DraineApJL1998,AliHaimoudMNRAS2009}. These grains must be small ($\la 10^{-9}$~m) in order
to spin fast and emit at $\approx30$~GHz: both PAHs \citep{DraineApJ1998} or nanosilicate
\citep{HoangApJ2016} have been proposed. Alternative explanations have AME produced by microwave enhancements 
in the dust emission cross section, due to magnetic inclusions in grains \citep{DraineApJ1999} or 
to their amorphous structure \citep{JonesA&A2009,NashimotoApJL2020}. Yet, there is no definitive
evidence for the effective emission mechanism: some works support the spinning-dust hypothesis
\citep{YsardA&A2010,HarperMNRAS2015,BellPASJ2019} while others find a fainter correlation of
AME with PAH emission \citep{HensleyApJ2016}.

Beyond the MW AME has been observed only in a handful of galaxies. It has been seen in a single 
star-forming  region in NGC~6946 \citep{MurphyApJL2010} and in a compact radio source in NGC~4725 
\citep{MurphyApJ2018}.
So far, the most convincing case for the detection of AME in the global Spectral Energy Distribution 
(SED) of a galaxy is that of M~31, the Andromeda galaxy. \citet{PlanckA&A2015} found that AME contributes 
to $\approx$35\% of the flux density at 30~GHz, though this was a marginal detection.
\citet{BattistelliApJL2019} claimed a stronger detection and an AME contribution of 75\%; 
they obtained it after constraining the synchrotron and thermal components at lower frequency
(6.6~GHz; see also \citealt{FatigoniA&A2021}), using observations from the 64-m Sardinia 
Radio Telescope \citep[SRT;][]{PrandoniA&A2017}. Much less prominent is AME in the Large and Small 
Magellanic Clouds (LMC and SMC, respectively), where it is found to contribute to 
$\approx$10\% of the SED at 22.8~GHz (PLA16).
In these objects (but also in M~31) a careful subtraction 
of the contribution from the Cosmic Microwave Background (CMB)
is needed. In fact, analysis based on previous
CMB templates claimed no detection for the LMC, and a stronger AME component for the SMC \citep{PlanckEarlyXVII}.
The search for AME in other galaxies proved unsuccessful (\citealt{PeelMNRAS2011}, hereafter P11; \citealt{TibbsMNRAS2018}; other unpublished non-detections are also reported by \citealt{DickinsonNAR2018}).

In this paper, we report on our attempt at detecting AME
in four nearby spiral galaxies, using SRT K-band 
(18 - 26.5 GHz) observations.
The sample, observations and image processing are described 
in Sect.~\ref{sec:sample}; photometry and SED fitting are presented in Sect.~\ref{sec:results}; the upper limits
we find for AME are discussed together with other observations in the literature in Sect.~\ref{sec:emissivity}; in 
Sect.~\ref{sec:conclusions} we draw our conclusions.

\section{Sample, observations and data reduction}
\label{sec:sample}

\begin{table}
\caption{Observing log.}              
\label{tab:log}      
\centering                                      
\begin{tabular}{lccccc}          
\hline\hline                        
Date & Band & $\tau$    & Target & N. of maps\\
     & GHz  &           &        & along RA+Dec\\
\hline                                   
03/01/21  & 24.6 & 0.18 & NGC~3627  & 11+11 \\
04/01/21  & 18.6 & 0.05 & NGC~4254  &  7+7  \\
05/01/21  & 18.6 & 0.05 & NGC~3627  &  7+6  \\
02/02/21  & 24.6 & 0.25 & NGC~4254  & 12+10 \\
05/02/21  & 24.6 & 0.16 & NGC~3627  &  9+9  \\
          &      &      & NGC~4254  &  4+3  \\
22/03/21  & 24.6 & 0.19 & NGC~4736  &  7+6  \\ 
          &      &      & NGC~5055  &  7+7  \\ 
31/03/21  & 18.6 & 0.08 & NGC~3627  &  1+1 \\
          &      &      & NGC~4736  &  7+7  \\
          &      &      & NGC~5055  &  7+7  \\
03/04/21  & 24.6 & 0.30 & NGC~5055  &  6+6 \\
\hline                                             
\end{tabular}
\end{table}

Our sample consists of four nearby spirals: NGC~3627 (M~66), NGC~4254 (M~99), NGC~4736 (M~94) and NGC~5055 (M~63). The objects were chosen for a pilot project
to extend the SED knowledge to the poorly known 
15-300 GHz frequency range \citep{BianchiProc2021}. In fact, an excellent
coverage is available in the UV-to-submm (from the DustPedia database; \citealt{DaviesPASP2017}) and in 
the mid radio-continuum \citep[][hereafter, T17]{TabatabaeiApJ2017}; furthermore, the galaxies are among the targets currently being mapped at millimeter wavelengths with the IRAM 30-m telescope and NIKA2 camera, within the IMEGIN program \citep{KatsioliProc2021,EjlaliProc2021}.

Observations at SRT were carried out during program 43-20 (January 2021) and related DDT programs 1-21 (February 2021) and 7-21 (March-April 2021).
We used the K-band 7-feed spectro-polarimetric receiver \citep{OrfeiIEEE2010}, coupled with the SArdinia Roach2-based Digital Architecture for Radio Astronomy back end (SARDARA; \citealt{MelisJAInst2018}). We observed two 1.2 GHz bands centered at 18.6 and 24.6 GHz (HPBW = 57$\arcsec$ and 45$\arcsec$, respectively). The bands were covered by 1.46 MHz frequency channels in full Stokes mode. Observations were done with an on-the-fly scheme over an area of $16'\times16'$, centered on each galaxy. In order to remove scan noise efficiently, several maps were taken along orthogonal RA and Dec directions, with scan speed 2$'$/s and scan separation HPBW/3 (each map lasting about 20 minutes). 
The observing log is given in Table~\ref{tab:log}.

The spectral cubes were reduced using the proprietary Single-dish Spectral-polarimetry Software (SCUBE; \citealt{MurgiaMNRAS2016}). Here we only describe the procedure used for total-intensity data, 
as no polarization was detected.
The flux density was brought to the scale of \citet{PerleyApJS2017} using the calibrators 3C\,147 and 3C\,286. The two sources were used also as band pass calibrators. 
We corrected the data to compensate for the atmosphere opacity that was derived by mean of sky dips performed during each observing session. We modeled the observed $T_{\rm sys}$ trends with
the airmass model to obtain the zenithal opacity, $\tau$ (see Table~\ref{tab:log}). 
We evaluated the consistency of the calibration by cross-checking the flux density of 3C\,147 against that of 3C\,286.
Using 3C\,147 as primary calibrator, and considering measurements in different weather conditions, we found that on average the observed 3C\,286 flux density is about $\pm 10$\% the expected value at both 18.6 and 24.6\,GHz. We assume the dispersion of the reproducibility of 3C\,286 flux density as an estimate of the systematic uncertainties in our calibration procedure due to errors in pointing, opacity correction, atmospheric fluctuations removal, etc. This systematic error is consistent with what found by \citet{LoiMNRAS2020} using a much larger data set.


We removed the baseline scan-by-scan by fitting a 2$^\mathrm{nd}$ order polynomial to the "cold-sky" regions around each target. We masked the target and removed all the surrounding point sources (from the 1.4\,GHz catalog by \citealt{CondonAJ1998}). In this way, we removed the contributions from the receiver noise, the atmospheric emission, and the large-scale foreground sky emission and we retained the target emission only, on scales smaller that the mask size. We then stack the data from all feeds to reduce the noise level. We combined the RA and Dec scans by mixing their stationary wavelet transform (SWT) coefficients (see \citealt{MurgiaMNRAS2016}). The SWT stacking, conceptually similar to the weighted Fourier merging \citep{EmersonA&A1988}, is very effective in isolating and removing the noise oriented along the scan direction. Finally, we average all spectral channels to increase the signal-to-noise ratio and produce maps with scale 15$\arcsec$/pixel.

\section{Results}
\label{sec:results}
\begin{figure*}
    \centering
    \includegraphics[width=\hsize]{"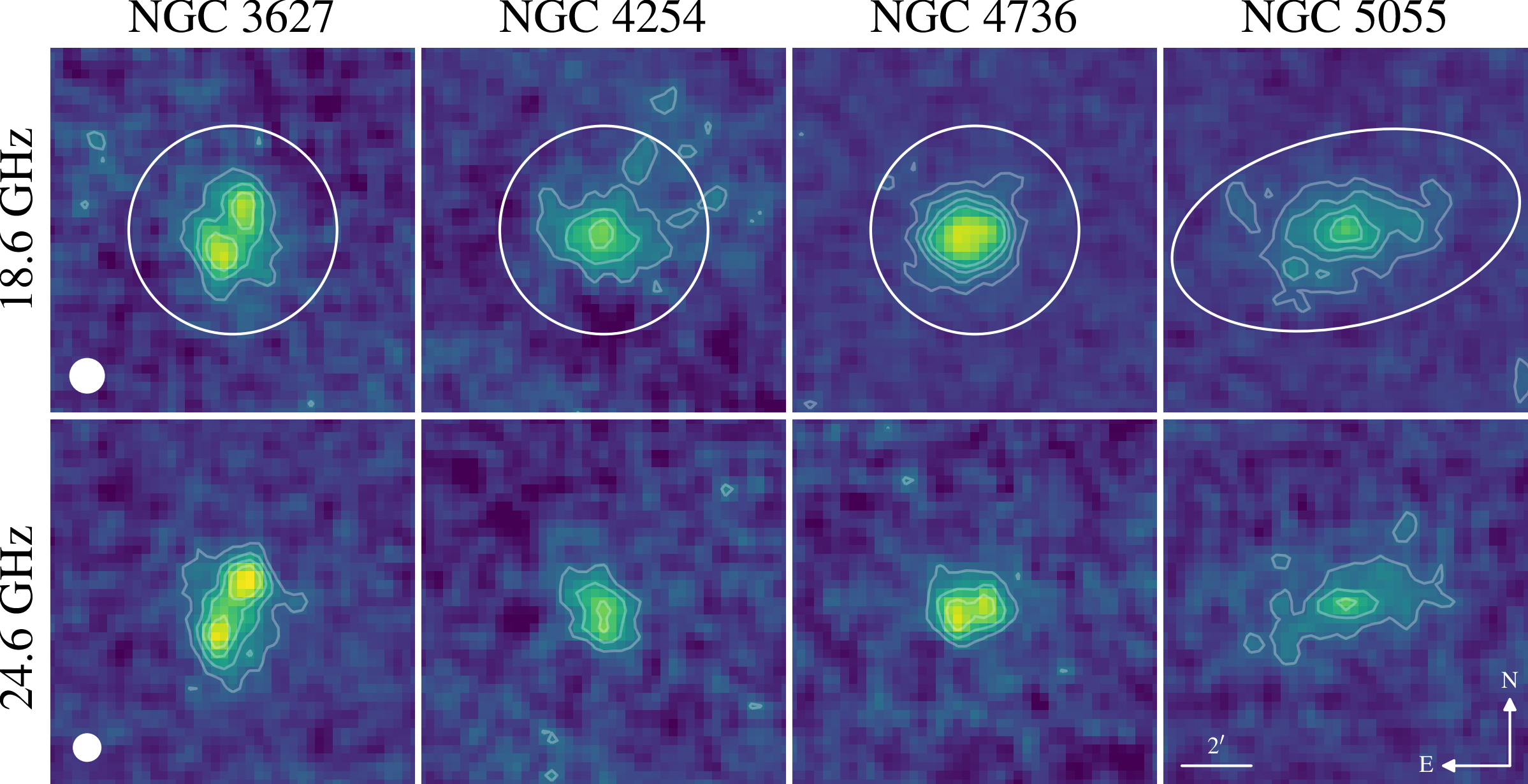"}
    \caption{
    Total intensity maps with contours at 3, 6, 9 and 12 $\times$ the sky $r.m.s.$. At 18.6~GHz, 
    $r.m.s. =$  1.2, 1.1, 0.6, 0.7 mJy beam$^{-1}$, for NGC~3627, NGC~4254, NGC~4736, NGC~5055, respectively; 
    at 24.6~GHz, $r.m.s. =$ 0.7, 0.8, 0.9, 0.7 mJy beam$^{-1}$. Photometry apertures and HPBWs are  also shown.
    }
    \label{fig:mapall}
\end{figure*}

\begin{figure*}
    \centering
    \includegraphics[width=\hsize]{"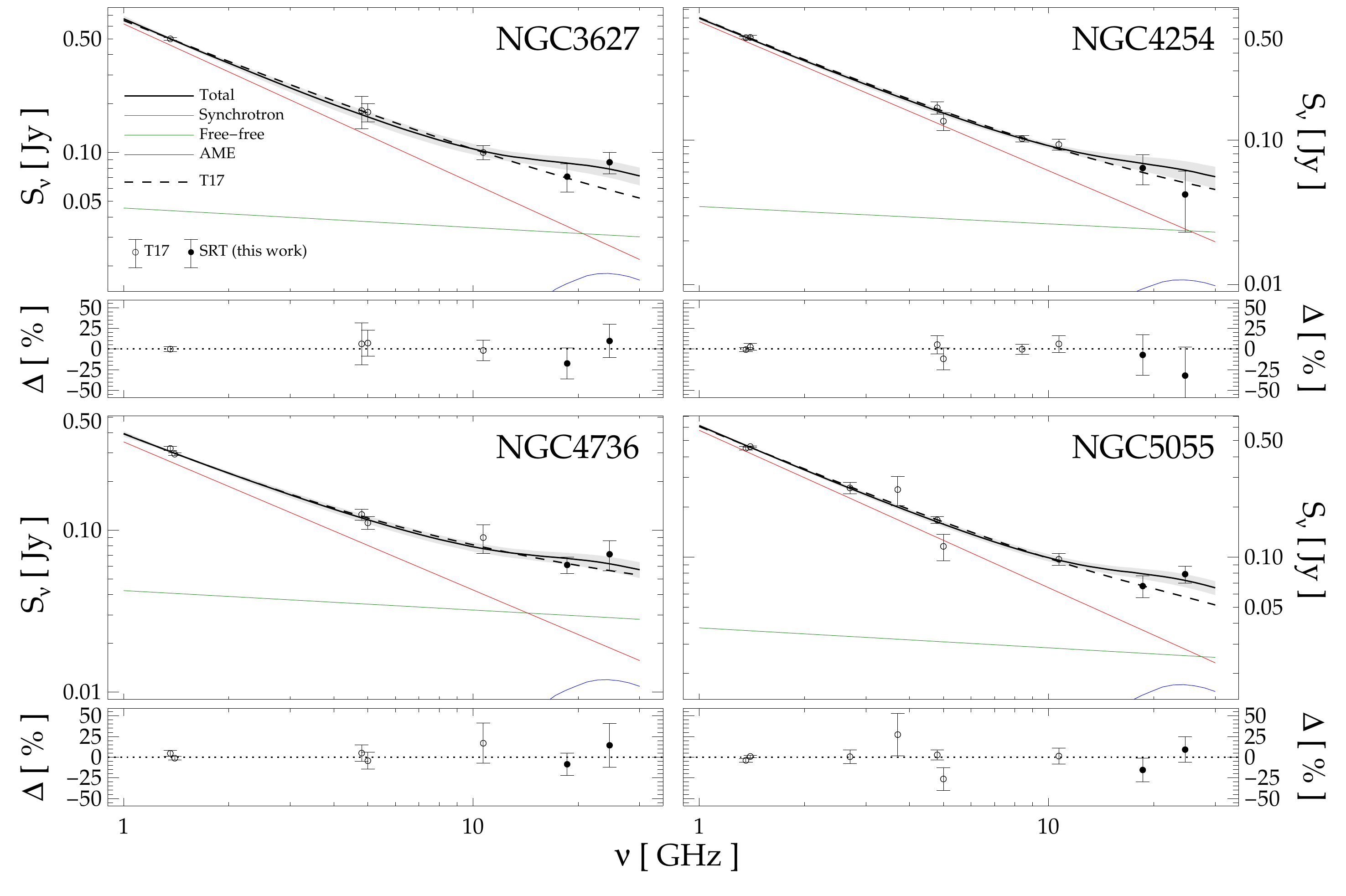"}
    \caption{
    Global SEDs and fits, with their uncertainty. We also show each SED component and the T17 fit. 
    Under each SED, we show the residuals $\Delta$=data-model (in percentage of the model).
    }
    \label{fig:sedall}
\end{figure*}

The total intensity maps of the four targets at 18.6 and 24.6~GHz are shown in Fig.~\ref{fig:mapall}. 
In all cases we detected diffuse emission, resembling the aspect, despite the lower resolution, 
observed in the far-infrared/submillimeter (FIR-submm) or in the radio (see Sect.~\ref{app:apertures} and Fig.~\ref{fig:mapcheck}). In particular, the emission in NGC~3627 peaks at both ends where the central bar meets the spiral arms, and parts of the ring of NGC~4736 are visible in the 24.6 GHz image.

In Fig.~\ref{fig:mapall} we also show the sky apertures used to derive 
the total flux density $S_\nu$. Due to the small size of our maps,
the apertures are taken to be small but sufficiently wide to enclose 
all the emission that is deemed to come from the galaxies, in both the 18.6 and 24.6 GHz maps. 
The same apertures were placed on a few positions outside of the targets to estimate the residual 
background and its fluctuations \citep[see, e.g., ][]{AuldMNRAS2013,ClarkA&A2018}. This procedure
gave results consistent with the more traditional sky estimate from an annulus around the source.
In Table~\ref{tab:observations} we give $S_\nu$ and its uncertainty, to which we added, in quadrature, 
the systematic error discussed in Sect.~\ref{sec:sample}.

\begin{table}
\caption{Aperture definition and SRT photometry.}
    \label{tab:observations}
    \centering
    \begin{tabular}{lcrcc}
   \hline \hline
        &\multicolumn{2}{c}{Apertures}& $S_\mathrm{18.6~GHz}$ & $S_\mathrm{24.6~GHz}$\\
        &Size&P.A.& mJy & mJy \\ \hline
NGC~3627 & $6\arcmin\times6\arcmin$& 0$^\circ$& $71\pm14$    & $87\pm13$    \\
NGC~4254 & $6\arcmin\times6\arcmin$& 0$^\circ$& $64\pm15$    & $42\pm19$   \\
NGC~4736 & $6\arcmin\times6\arcmin$& 0$^\circ$& $61\pm7$    & $71\pm15$    \\
NGC~5055 & $10.2\arcmin\times5.5\arcmin$& 103$^\circ$& $67\pm10$    & $79\pm9$    \\ \hline
\end{tabular}
\end{table}

\begin{table*}[!ht]
\caption{
Fit parameters, relative contribution of the various components at 5 and 24.6 GHz and upper limits for $S_\mathrm{24.6~GHz}^\mathrm{AME}$.
}
    \label{tab:fits}
    \centering
    \begin{tabular}{lcccc|cc|ccc|c}
    \hline \hline
         & $\alpha$  & $S^\mathrm{sy}_\mathrm{5~GHz}$ & $S^\mathrm{ff}_\mathrm{5~GHz}$&  $S_\mathrm{24.6~GHz}^\mathrm{AME}$ &
         $S^\mathrm{sy}_\mathrm{5~GHz}$ & $S^\mathrm{ff}_\mathrm{5~GHz}$ &
         $S^\mathrm{sy}_\mathrm{24.6~GHz}$ & $S^\mathrm{ff}_\mathrm{24.6~GHz}$ & $S_\mathrm{24.6~GHz}^\mathrm{AME}$ &
         $S_\mathrm{24.6~GHz}^\mathrm{AME~u. l.}$\\
         &           & mJy & mJy & mJy & \% & \% & \% & \% & \% & mJy \\ \hline
NGC~3627 &  0.98$^{+0.20}_{-0.12}$ & 127$^{+30}_{-37}$ & 38$^{+29}_{-26}$ & 18$^{+16}_{-12}$ & 77 & 23 & 35 & 41 & 24 & $<44$ \\
NGC~4254 &  1.03$^{+0.16}_{-0.09}$ & 125$^{+22}_{-29}$ & 29$^{+23}_{-19}$ & 11$^{+12}_{-8}$ & 81 & 19 & 41 & 41 & 18 & $<32$ \\
NGC~4736 &  0.91$^{+0.21}_{-0.13}$ & 81$^{+23}_{-25}$ & 35$^{+23}_{-22}$ & 12$^{+11}_{-8}$  & 70 & 30 & 32 & 48 & 20 & $<30$\\
NGC~5055 &  0.94$^{+0.12}_{-0.08}$ & 126$^{+21}_{-25}$ & 31$^{+23}_{-21}$ & 17$^{+12}_{-11}$ & 80 & 20 & 40 & 36 & 24 & $<36$\\ \hline
\end{tabular}
\end{table*}

In Fig.~\ref{fig:sedall} we show the SEDs of the four galaxies in the 1-30 GHz range, 
using the SRT photometry and the mid-radio continuum data from T17. Ideally, one should
use flux densities obtained by integrating maps over the same aperture at all frequencies. 
Instead, the flux densities in the compilation of T17 refer to apertures based on the 
optical size of each galaxy, which are larger than what we used above. Nevertheless,
we find that our choice does not alter significantly the estimate of the flux density
at SRT frequencies (tests are described in Sect.~\ref{app:apertures}).
We model the SED as the sum of three continuum components: i) synchrotron radiation,
dominating in the upper end of the range, which -in the optically thin regime- can be described by a power-law;
ii) free-free, emission, almost flat with frequency, emerging over synchrotron as the frequency increases; iii)
AME, expected to contribute at the highest frequencies considered here. It is:
\begin{equation}
    S_\nu = 
    S^\mathrm{sy}_\mathrm{5~GHz}\times \left(\frac{\nu}{\mathrm{5~GHz}}\right)^{\alpha} +
    S^\mathrm{ff}_\mathrm{5~GHz}\times \left(\frac{\nu}{\mathrm{5~GHz}}\right)^{-0.12} +
    S^\mathrm{AME}_\nu,
    \label{eq:model}
\end{equation}
where $\alpha$ is the synchrotron spectral index, $S^\mathrm{sy}_\mathrm{5~GHz}$ and $S^\mathrm{ff}_\mathrm{5~GHz}$ are the amplitudes of the 
synchrotron and free-free component at 5~GHz, respectively, and the weak frequency 
dependence of the free-free continuum is taken from \citet{DraineBook2011}. 
For the AME component, we adopted the spectral template of \citet{BattistelliApJL2019}, derived 
by averaging theoretical predictions over a variety of ISM environments, using the {\tt spdust.2} code 
\citep{AliHaimoudMNRAS2009,SilsbeeMNRAS2011} and neglecting the most extreme 
cases presented in Fig.~2 of \citet{DickinsonNAR2018}.
We scale the template on its amplitude at 24.6~GHz, $S_\mathrm{24.6~GHz}^\mathrm{AME}$ and, as in 
\citet{BattistelliApJL2019}, we keep its peak fixed (it is located at $\nu \approx 25$ GHz). 
We do not include thermal emission from dust, negligible in the spectral range we considered (see, e.g., P11).

We fit Eq.~\ref{eq:model} to the data using the {\tt emcee} Python implementation (version 3.0) 
of the Goodman \& Weare’s Affine Invariant Markov chain Monte Carlo (MCMC) Ensemble sampler \citep{ForemanMackeyPASP2013}. We adopted positive flat priors for $\alpha$, 
$S^\mathrm{sy}_\mathrm{5~GHz}$, $S^\mathrm{ff}_\mathrm{5~GHz}$,  $S_\mathrm{24.6~GHz}^\mathrm{AME}$
and obtained their posterior probability distribution function (PDF) by sampling the likelihood function, 
conditional on the data. 
We used 100 walkers, initially distributed around the T17 fit (assuming negligible AME), and run a chain 
of $2\times 10^4$ steps. The PDFs were obtained after deriving the integrated autocorrelation time and 
neglecting a few times this number of steps to remove the {\em burn-in} phase from the chain. We tested 
on the convergency of the results by varying the initial conditions (setting higher values for AME), the 
number of walkers and samples. The fits and  uncertainties are shown in Fig.~\ref{fig:sedall},
while in Table~\ref{tab:fits} we present the estimate and uncertainty of each parameter,
derived from the median and 0.16 and 0.84 percentiles
(see Fig.~\ref{fig:corner} for the Bayesian corner plots).

\begin{figure}
    \centering
    \includegraphics[width=\hsize]{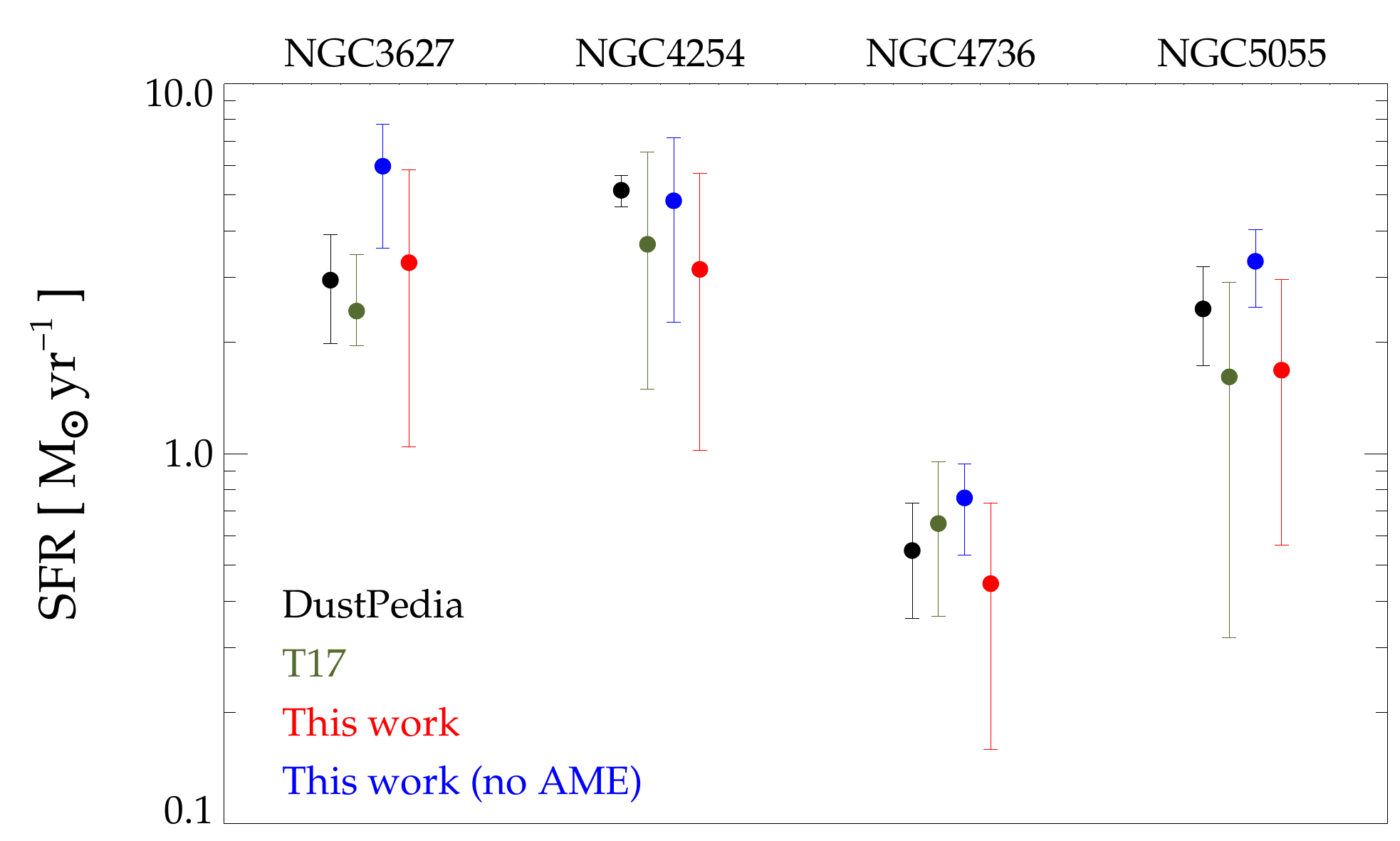}
    \caption{Comparison between SFRs from DustPedia and from $S_\mathrm{5 GHz}^\mathrm{ff}$ (see text for details).}
    \label{fig:ff_sfr}
\end{figure}

The only parameters estimated at high significance are those for the synchrotron component:
in particular, our $\alpha$ estimates are consistent within the errors with those of T17. 
Much more uncertain is the estimate of the free-free component: it is degenerate with
that of synchrotron at lower frequencies, contributing to 20-30 \% of the continuum flux density at 5 GHz (see Table~\ref{tab:fits}); it is also degenerate with the AME component, as the largest contribution to the SED of both emissions 
insists on the same spectral window, for $\nu \ga 15$~GHz. Because of this, we expected our estimates to 
be more uncertain than those in T17, that do not include an AME component and use only data with 
$\nu\la 10$~GHz. 
We compare the results from the two fits in Fig.~\ref{fig:ff_sfr}, where we have converted 
$S_\mathrm{5~GHz}^\mathrm{ff}$ into the SFR, since the free-free emission is proportional 
to the amount of free electrons, and thus to the photoionizing rate (\citealt{MurphyApJ2011}; T17). 
Despite the larger uncertainties (but not for NGC~5055), our results are still compatible with T17.
Both estimates are also broadly consistent with the SFRs derived by \citet{NersesianA&A2019} from 
the analysis of the whole DustPedia UV-to-submm SED, thus independently of observations in the radio. 
As shown by Fig.~\ref{fig:ff_sfr}, the advantage of
deriving the SFR from the radio thermal 
radiation, in a spectral range unaffected by dust extinction,
is compounded by the larger uncertainties due to the fit degeneracies. Data at frequencies 
above 25~GHz is needed to solve these degeneracies and obtain more precise determinations of SFR 
from the radio continuum and of the presence of AME.

All three components contribute to the continuum at 24.6~GHz (see Table~\ref{tab:fits}):
the synchrotron and free-free fractions are about the same, $\sim 40\%$, while the AME fraction
is about half that value, $\sim 20\%$. However, in no case the fitted $S_\mathrm{24.6~GHz}^\mathrm{AME}$ 
is significant. Thus, in the following discussion we use upper limits for 
$S_\mathrm{24.6~GHz}^\mathrm{AME}$, derived from the 0.95 percentile of the PDF.

Within this work, we also tried to fit the SED without including the AME component 
in Eq.~\ref{eq:model}: upper limits on $S^\mathrm{AME}_\mathrm{24.6~GHz}$ were then
obtained from the residuals between observations and model (as done in P11). In this case we still 
obtained plausible fits (not shown) by steepening the synchrotron spectrum and raising 
the free-free contribution: with respect to fits with AME included,
$S_\mathrm{5~GHz}^\mathrm{ff}$ values are larger by a factor $\sim$2 (see the 
corresponding SFRs in Fig.~\ref{fig:ff_sfr}) while the AME upper limits becomes smaller.
In fact, as pointed out by P11, this approach might result in a bias towards a 
lower AME estimate. Nevertheless, either using the $S^\mathrm{AME}_\mathrm{24.6~GHz}$ upper limits
from the full fit or those from the residuals to a no-AME fit has no consequence 
for the rest of the discussion in this work.

\section{AME emissivity}
\label{sec:emissivity}

In the MW, AME has often been studied in correlation with dust emission at 100~$\mu$m (3~THz) \citep{DickinsonNAR2018}. For Galactic latitudes $|b|>10^\circ$, PLA16 find $S^{AME}_{\mathrm{22.8~GHz}}/S_\mathrm{3~THz}=3.4\pm0.3 \times 10^{-4}$. 
Usually, the MW estimate has been compared with observations in other galaxies: while
\citet{BattistelliApJL2019} derive in M31 a value for $S^{AME}_{\mathrm{30~GHz}}/S_\mathrm{3~THz}$ 
consistent with the MW, P11 and \citet{TibbsMNRAS2018} find smaller ratios for their targets; 
the same is true for our targets.
For simplicity, we use the same notation for the MW (and Magellanic Clouds), where the ratio is 
derived by correlating surface brightness maps, and the rest of the galaxies we discuss, where it 
is a ratio between flux densities (equivalent to a ratio between galaxy-averaged surface brightnesses). 
Also, the ratios are obtained at different frequencies: nevertheless they are not expected to change 
much in the 20-30~GHz range (by $\la10$\% for the template of \citealt{BattistelliApJL2019}).
Thus, we neglect any frequency dependence and use for all objects the term $S^{AME}_{\mathrm{30~GHz}}/S_\mathrm{3~THz}$, also 
when referring to the SRT 24.6~GHz estimates (see Table~\ref{tab:other}).

\citet{TibbsAinA2012} questioned the validity of $S^{AME}_{\mathrm{30~GHz}}/S_\mathrm{3~THz}$ as 
an indicator of the AME emissivity: in fact, the 100~$\mu$m flux density - near the peak of dust emission - 
depends strongly on the grain temperature; instead, under the spinning-grain hypothesis, 
the low intensity radiation fields responsible for diffuse dust emission have little 
influence on AME, at least for low density environments \citep{AliHaimoudMNRAS2009,YsardA&A2011}. 
A better characterization for AME is to derive its emissivity, i.e.\ its surface brightness $I^\mathrm{AME}_{\nu}$ normalized to a tracer of the amount of underlying emitting material. Under the assumption that AME is due to dust, we can write
\begin{equation}
    \epsilon^\mathrm{AME}_{\mathrm{30~GHz}} =
\frac{I^\mathrm{AME}_{\mathrm{30~GHz}}}{\Sigma_\mathrm{d}} =
\frac{S^\mathrm{AME}_{\mathrm{30~GHz}}}{S_{\mathrm{3~THz}}} \kappa^\mathrm{d}_\mathrm{3~THz} B_\mathrm{3~THz}(T_\mathrm{d}),
\label{eq:emy2}
\end{equation}
where $\Sigma_\mathrm{d}$ is the dust mass surface density, $B_\mathrm{3~THz}(T_\mathrm{d})$ the Planck 
function for the dust temperature $T_\mathrm{d}$ and $\kappa^\mathrm{d}_\mathrm{3~THz}$ the dust absorption cross-section\footnote{
Equivalently, Eq.~\ref{eq:emy2} can be written as
\begin{equation}
\epsilon^\mathrm{AME}_{\mathrm{30~GHz}} =
\frac{S^\mathrm{AME}_{\mathrm{30~GHz}} }{M_\mathrm{d}/D^2},
\label{eq:emy2a}
\end{equation}
where  $M_\mathrm{d}$ is the galaxy's dust mass and
$D$ is the distance used for its determination (Eq.~\ref{eq:emy2a}, as Eq.~\ref{eq:emy2}, being eventually independent on $D$; see also \citealt{BianchiA&A2019}).
}. 
For our targets, temperatures (and dust masses) were obtained from single-temperature modified 
blackbody (MBB) fits to FIR/summ SEDs by \citet{NersesianA&A2019}, after assuming the average 
properties of MW diffuse dust from The Heterogeneous dust Evolution Model for Interstellar Solids (THEMIS; \citealt{JonesA&A2017}): in particular, THEMIS dust is characterized by an effective absorption 
cross section that can be described by $\kappa^\mathrm{d}_\nu$ / (pc$^2$/M$_\odot$) $ = 6.9\times 10^{-3} \times 
(\nu/\mathrm{3~THz})^{1.79}$ \citep{GallianoARA&A2018}.

\begin{figure}
    \centering
    \includegraphics[width=\hsize]{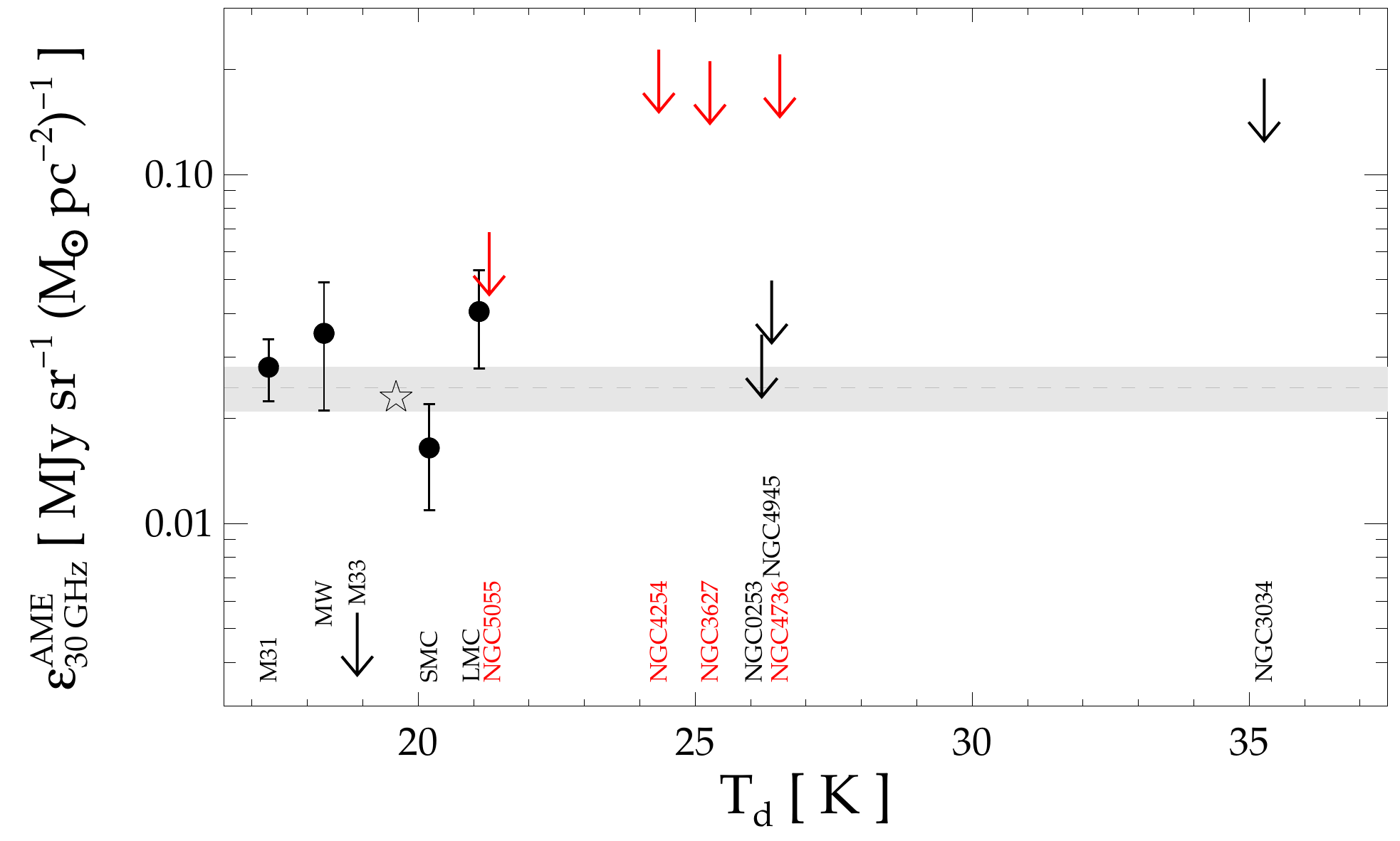}
    \caption{$\epsilon^\mathrm{AME}_{\mathrm{30~GHz}}$ vs $T_\mathrm{d}$. Red symbols are for SRT targets. 
    The shaded area shows the weighted average and standard deviation of
    $\epsilon^\mathrm{AME}_{\mathrm{30~GHz}}$; the star is the PLA16 estimate for the MW (see text for details).
    }
    \label{fig:emy1}
\end{figure}

The emissivity $\epsilon^\mathrm{AME}_{\mathrm{30~GHz}}$ is shown as a function of $T_\mathrm{d}$ in 
Fig.~\ref{fig:emy1}, where we also included the estimates for the MW and M~31,
those for the LMC and SMC (PLA16); and the upper limits for M~33 \citep{TibbsMNRAS2018} and  
NGC~253, NGC~3034 and NGC~4945 (P11). Seeking for uniformity in the estimate of dust parameters we 
performed MBB/THEMIS fits also for these objects (Table~\ref{tab:other}).
Almost all the upper limits are compatible with the estimates in the MW and M~31 and also the LMC and SMC 
values are not far off. Only for M~33 $\epsilon^\mathrm{AME}_{\mathrm{30~GHz}}$ is much lower, suggesting 
different properties of the 
interstellar medium (ISM) and/or the dust grains for this galaxy with respect to the MW and M~31.
After computing the spinning-grain emission for the THEMIS dust distribution, most of the spread in Fig.~\ref{fig:emy1} can be reconciled with variations in the ISM density and radiation field 
intensity (Ysard et al. in preparation).

The weighted average of the four determinations is $\epsilon^\mathrm{AME}_{\mathrm{30~GHz}} =
2.4\pm0.4 \times 10^{-2}$ MJy sr$^{-1}$ (M$_\odot$ pc$^{-2}$)$^{-1}$ (shaded area in Fig.~\ref{fig:emy1}).
For the dust-to-gas mass ratio of the THEMIS model (0.0074), this converts to
$1.4\pm0.2 \times 10^{-18}$ Jy sr$^{-1}$ (H cm$^{-2}$)$^{-1}$ (in units of H column density).
A possible caveat here can be due to the uncertainties inherent to the determination of the dust column 
density and to the necessity of using coherent methods for all the objects. 
For example, 
if we convert the average emissivity to units of surface brightness per dust optical 
depth at 850~$\mu$m (it is $S^\mathrm{AME}_\mathrm{30~GHz}/\tau^\mathrm{d}_\mathrm{353~GHz} = 
\epsilon^\mathrm{AME}_{\mathrm{30~GHz}}/\kappa^d_\mathrm{353~GHz}$) and express it in 
temperature (at 22.8~GHz) we obtain $10\pm2$~K, a value close to what PLA16 derive - but for the MW only.
The difference might reside in the different reference frequency 
for estimating the dust surface density, and in the
derivation of the dust temperature from the observed SED (using a fixed 
dust model, in our case, vs a fit of the opacity cross section dependence on frequency in the Planck analysis, 
resulting in $T_\mathrm{d}=19.6$~K in PLA16; star in Fig.~\ref{fig:emy1}). The fixed
power-law expression for $\kappa^\mathrm{d}_\nu$ we adopted might not be the best choice for describing dust emission, when using a single MBB; nevertheless it proved reliable in the estimate of the dust mass (and surface density) under different heating conditions, which is important for the current analysis
(see, e.g. \citealt{NersesianA&A2019}). Using the average $\epsilon^\mathrm{AME}_{\mathrm{30~GHz}}$, we 
can estimate $S^\mathrm{AME}_{\mathrm{30~GHz}}$ from $S_{\mathrm{3~THz}}$ and the fitted $T_\mathrm{d}$ 
(inverting Eq.~\ref{eq:emy2}) or from $M_\mathrm{d}$ (with Eq.~\ref{eq:emy2a}): we derive
$S^\mathrm{AME}_{\mathrm{30~GHz}}\approx 5$~mJy (15 mJy for NGC~5055), on the lower side of the 
spread of Table~\ref{tab:fits} but still compatible within the large uncertainties from the fit.

\begin{figure}
    \centering
    \includegraphics[width=\hsize]{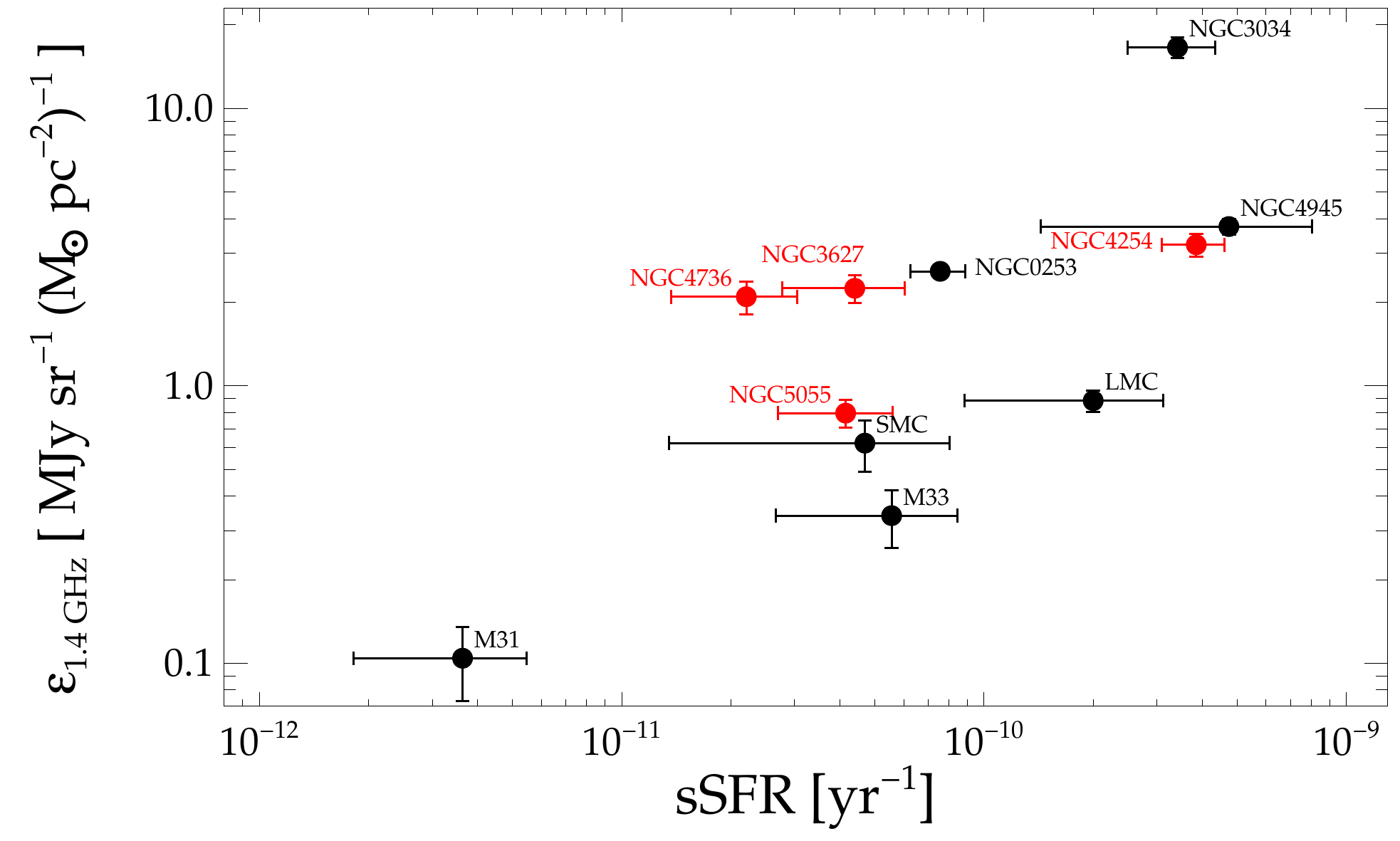}
    \caption{$\epsilon_\mathrm{1.4~GHz}$ vs specific star formation rate, $sSFR$. Red symbols are for SRT targets. See  Table~\ref{tab:other} for the source of $SFR$s and $M_\star$ used to derive $sSFR$.}
    \label{fig:sfr}
\end{figure}

In an attempt to understand what makes the detection of AME peculiar in M~31 (where $S^\mathrm{AME}_{\mathrm{30~GHz}}$ is
estimated to be about 35-70\% of the total flux density, 
while it is a smaller fraction for the rest of the sample) we defined the radio emissivity per 
dust column density, $\epsilon_{\mathrm{1.4~GHz}}$, analogous to 
$\epsilon^\mathrm{AME}_{\mathrm{30~GHz}}$ in Eq.~\ref{eq:emy2} and \ref{eq:emy2a}
(this quantity is equivalent to divide $S_\mathrm{1.4~GHz}$ by the {\em predicted} $S_\mathrm{30~GHz}^\mathrm{AME}$, if a single $\epsilon^\mathrm{AME}_{\mathrm{30~GHz}}$ 
is valid for all galaxies). In Fig.~\ref{fig:sfr} we plot $\epsilon_{\mathrm{1.4~GHz}}$
against the specific star formation rate, i.e.\ the star-formation rate per unit stellar 
mass, $sSFR = SFR / M_\star$, a physical quantity that shows the largest dichotomy between 
M~31, at the lowest value, and the rest of the sample. 
If we do not consider M~33, which apparently has different AME properties,
$\epsilon_\mathrm{1.4~GHz}$ for M~31 is almost an order of magnitude smaller than for 
the other galaxies considered here. Even though more observations (and AME detections) 
are needed to confirm this apparent trend, one might wonder if AME could be better observed 
in quiescent spirals in the final stages of their evolution, such as Andromeda, where 
the build-up of the dust mass has concluded, so that AME intensity is higher, and the 
radio continuum has a lower luminosity, thus allowing AME to emerge over the other SED 
components.

\section{Conclusions}
\label{sec:conclusions}

We have shown that the upper limits derived for AME in four galaxies observed at SRT 
(and those for three other objects in the literature) are consistent with 
the few AME detections, and in particular those for the MW and M~31, once a proper 
AME emissivity per dust surface density is defined.

The analysis presented here is only tentative, as it is based only on a few detections.
Also, it implicitly relies on the main assumption that AME does not change from object to
object, and that its spectrum peaks in between 20 to 30 GHz, as in the diffuse MW medium and
in M~31. We already found in the small sample we analyzed one object, M~33, whose AME emissivity
is substantially smaller than those for the other galaxies, possibly on account of
a reduced fraction of the small grains responsible for AME \citep[see, e.g.,][]{CalapaApJ2014,WilliamsMNRAS2019}. 
Also, the spinning-grain theory predicts 
(and some observations in the MW suggest) that the AME spectrum peaks at higher frequency
in higher density environments \citep[see, e.g., ][]{DickinsonNAR2018}.
Thus, different galaxies might have a different AME spectrum depending on the dominant ISM
conditions (and on their relative proportion).

Certainly, observations of the full SED beyond 30~GHz and up to the millimeter spectral range
could provide further opportunities for new detections and the characterization of AME. 
Furthermore, they will offer the possibility to solve the degeneracies in the continuum
decomposition, without which it will not be possible to fully exploit the use of free-free 
radiation as a dust-free tracer of the SFR. At the SRT, high frequency observations will 
become possible soon with the ongoing upgrade to new 
Q-band (33-50~GHz) and W-band (70-116~GHz) instruments \citep{GovoniProc2021}.

\begin{acknowledgements}
This paper is dedicated to the memory of Jonathan I. Davies,
to whom we are grateful, among many other things, for devising the
DustPedia and IMEGIN projects.
We thank E.\ S.\ Battistelli for useful comments and F. Radiconi for sharing with us the AME template used for the M31 analysis. 
SB and VC acknowledge support from the INAF main stream 2018 program
``Gas-DustPedia: A definitive view of the ISM in the Local Universe'', and from the grant PRIN MIUR 2017 - 20173ML3WW\_001.
The Sardinia Radio Telescope is funded by the Ministry of University and Research (MIUR), Italian Space Agency (ASI), and the Autonomous Region of Sardinia (RAS) and is operated as National Facility by the National Institute for Astrophysics (INAF). 
The Enhancement of the Sardinia Radio Telescope (SRT) for the study of the Universe at high radio frequencies is financially supported by the National Operative Program (Programma Operativo Nazionale - PON) of the Italian Ministry of University and Research "Research and Innovation 2014-2020", Notice D.D. 424 of 28/02/2018 for the granting of funding aimed at strengthening research infrastructures, in implementation of the Action II.1 – Project Proposals PIR01\_00010 and CIR01\_00010.

\end{acknowledgements}

\bibliographystyle{aa} 
\bibliography{DUST} 

\begin{appendix}

\section{Additional material}
\label{app:tables}

\subsection{Tests on aperture photometry}
\label{app:apertures}

 \begin{figure*}
    \centering
    \includegraphics[width=\hsize]{"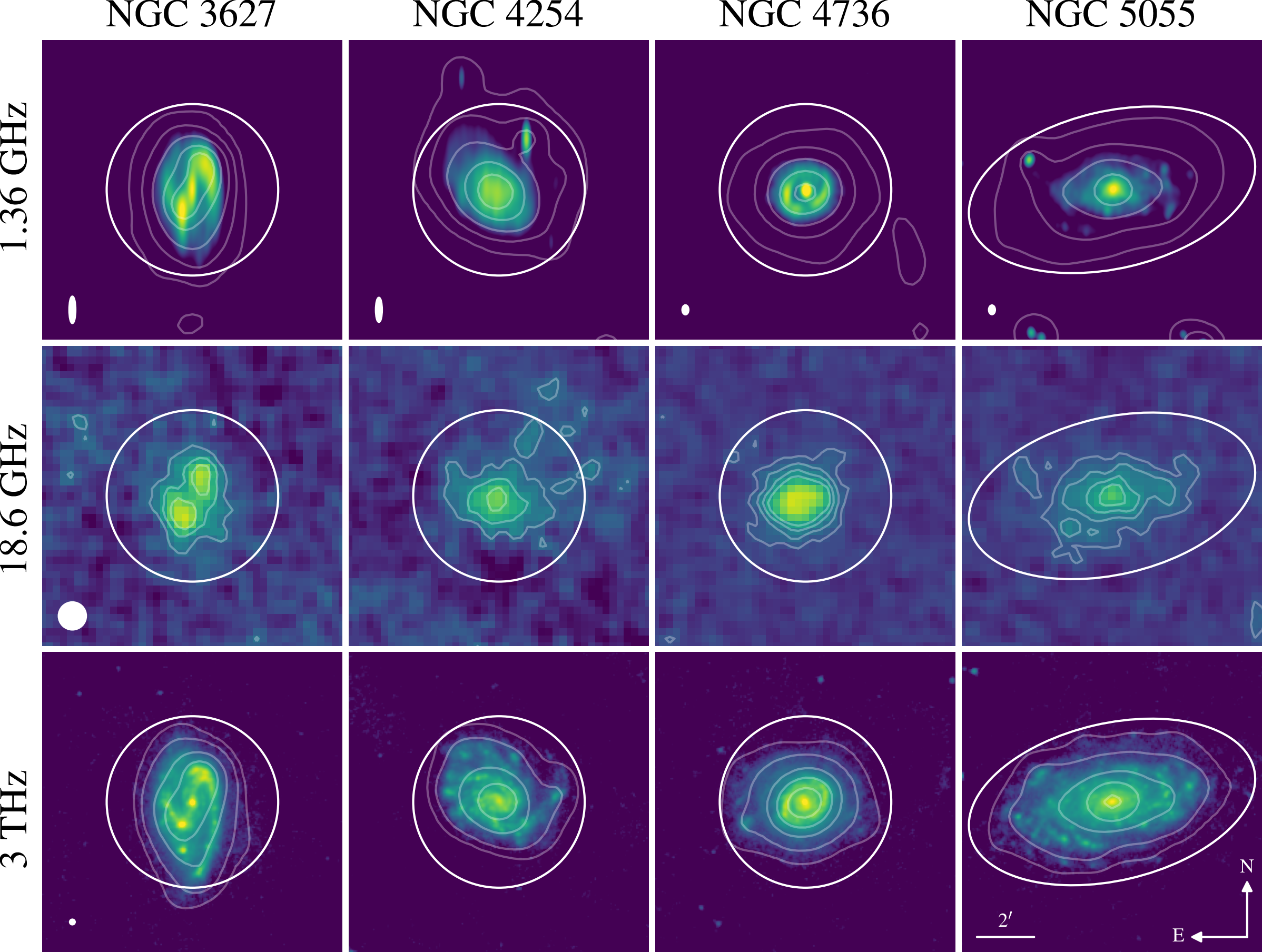"}
    \caption{Top panels: Westerbork Synthesis Radio Telescope images from \citet{BraunA&A2007}; the white ellipses show
    the synthesized beams; the contours show the same maps smoothed to the SRT 18.6~GHz resolution (HPBW=57$\arcsec$), at surface brightness levels of 0.025, 0.1, 0.25, 0.5, 0.75 MJy sr$^{-1}$.
    Bottom panels: {\em Herschel} 3~THz (100 $\mu$m) maps from DustPedia, with HPBW=10$\arcsec$ (white circle); the contours show the maps smoothed to the SRT resolution, at 5, 20, 100, 250, 500 MJy sr$^{-1}$.
    As a reference, the SRT images at 18.6~GHz are shown in the central panel (same as in Fig.~\ref{fig:mapall}).
    The apertures adopted in our work are also shown.
    }
    \label{fig:mapcheck}
\end{figure*}

Our apertures are smaller than those used to derive the flux densities in T17, 
which instead cover the optical size of each galaxy. For example, the 1.36~GHz flux densities 
listed in T17 have been obtained by \citet{BraunA&A2007} using circular apertures of 
size 10 to 15$\arcmin$, almost reaching the whole extent of the SRT maps. 
Since the original Westerbork Synthesis Radio Telescope images of \citet{BraunA&A2007}
are available from the NASA Extragalactic Database (NED), we used them to test
the impact of our smaller apertures. We follow the procedure of \citet{DaleApJ2012} 
to estimate the amount of flux that would go beyond the aperture because of beam smearing.
We assume that the 1.36~GHz image can be used as a higher-resolution proxy for the 
surface brightness spatial distribution at the SRT frequencies, and smooth it to
the SRT 18.6~GHz resolution, using Gaussian kernels. For each of our targets,
the full resolution image is shown in Fig.~\ref{fig:mapcheck}, with superimposed 
contours from the smoothed image. We found that the global flux densities of 
\citet{BraunA&A2007} are larger than those derived by integrating the 
smoothed maps on our apertures, but only by a moderate factor: 3 to 5\% for 
NGC~3627, NGC~4254 and NGC~4736. The total flux density at 1.36~GHz for NGC~5055 
is instead 12\% larger. However, this is due in part to point sources non related 
to the galaxy (some of which are visible in the southern edge of the map in 
Fig.~\ref{fig:mapcheck}): when these are removed, the difference reduces to 5\%.
Since the errors on SRT photometry are larger, we do not apply any correction.
We only note that, for NGC~5055, the additional flux density picked from the
background sources might affect also the other large-aperture flux densities quoted 
by T17. 
Thus, before SED fitting, one should lower those flux densities down by about 7\%. 
Even if this slightly changes the values of the fitted parameters, it has little effect
on the general results for this galaxy. Thus, we do not implement this correction either.

In this work we also made use of DustPedia 3~THz flux densities, obtained by 
integrating {\em Herschel} 100 $\mu$m images over apertures matched to contain 
all emission from the UV to the submm \citep{ClarkA&A2018}. We repeated the same 
exercise on the images from the DustPedia archive: the full resolution maps
are shown in the bottom panel of Fig.~\ref{fig:mapcheck}, with superimposed 
contours from the same maps smoothed to the SRT resolution (using the convolution 
kernels of \citealt{AnianoPASP2011}). This time the flux densities obtained over 
the larger aperture are almost the same, for NGC~3627 and NGC~4736. 
For NGC~4254 and NGC~5055, instead, the DustPedia flux densities are {\em smaller}
than those derived on the apertures defined here. Nevertheless, the difference
is still compatible with the photometric uncertainties, highlighting the fact that
apertures much larger than the apparent size of the objects, though functional in 
detecting all of the emission, can also result in larger uncertainties,
due to fluctuations in the sky background and unremoved sources other than 
the one of interest.

\subsection{Details on SED fitting}
\label{app:fitdetails}

\begin{figure*}
    \centering
    \includegraphics[width=9.2cm]{"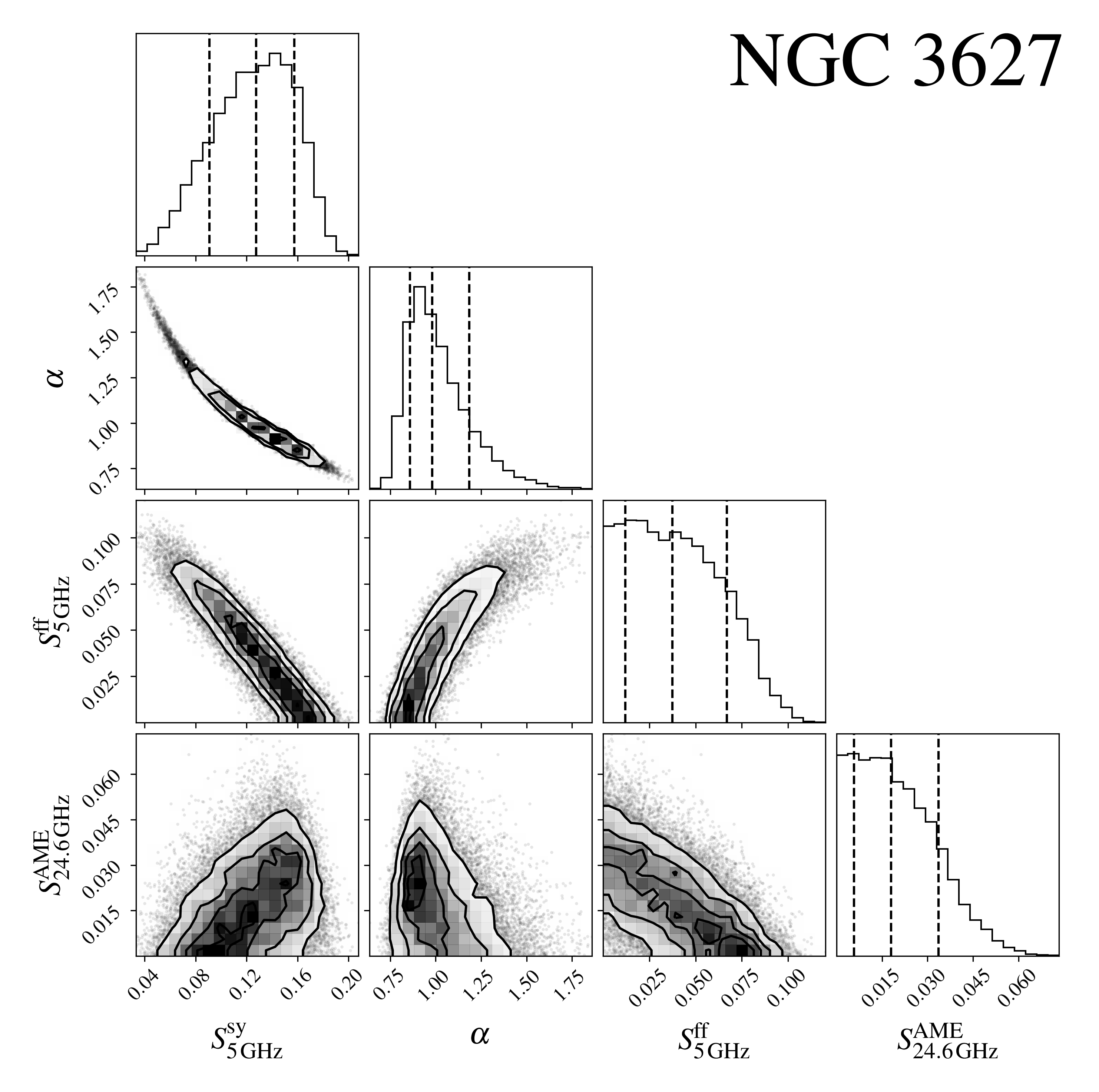"}\includegraphics[width=9.2cm]{"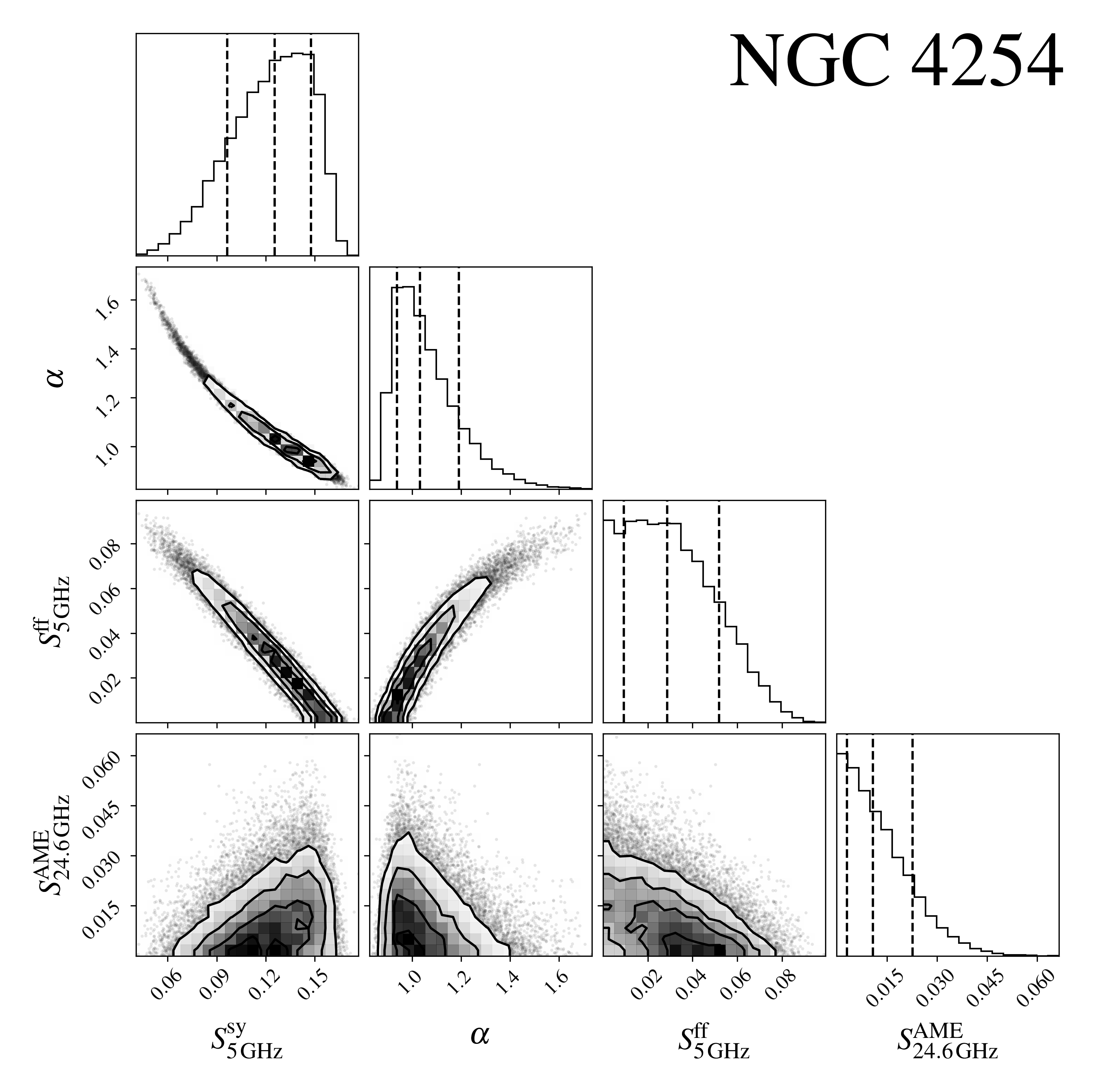"}
    \includegraphics[width=9.2cm]{"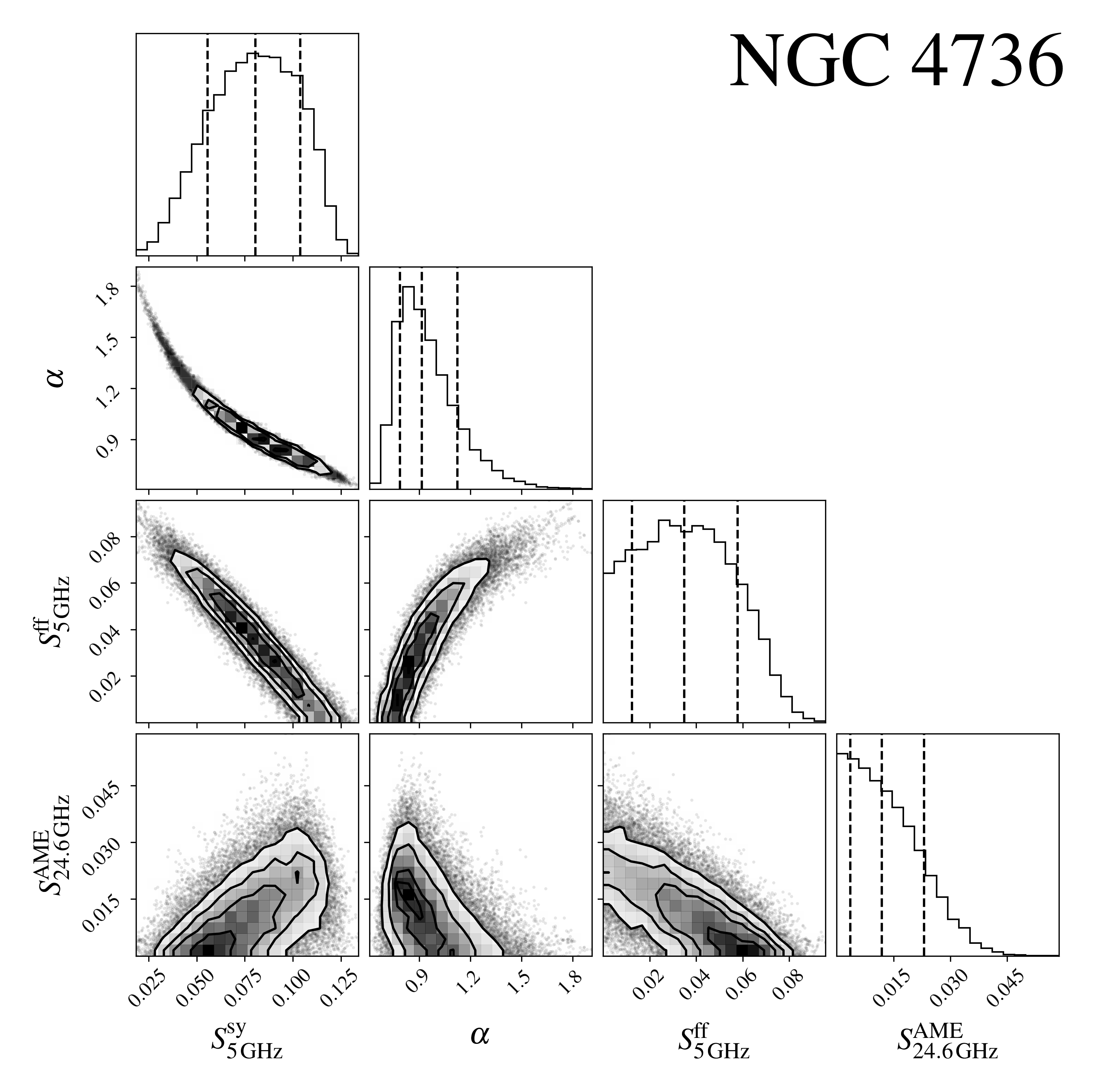"}\includegraphics[width=9.2cm]{"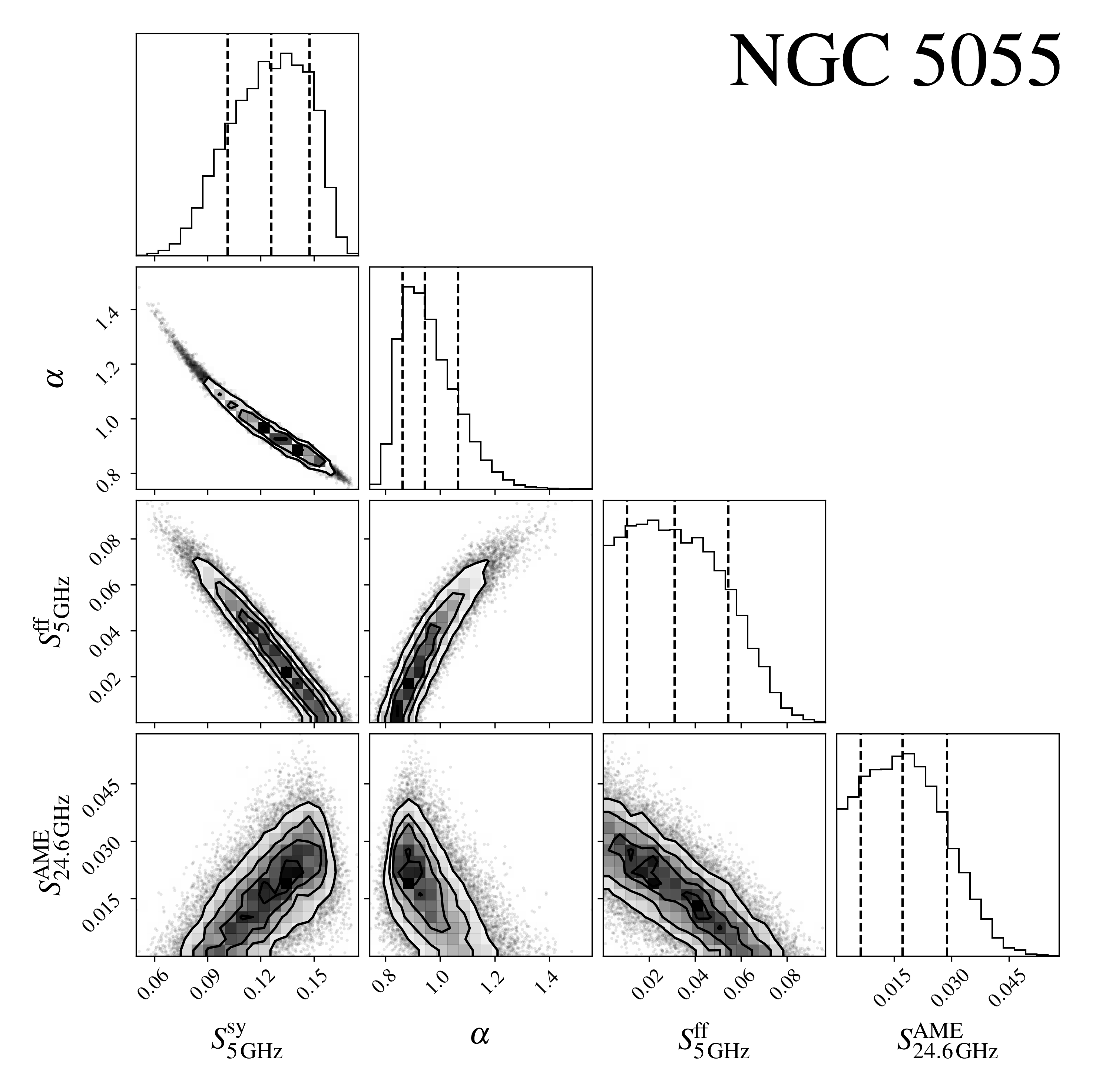"}
    \caption{
    Corner plots for the parameters describing the SED. Flux densities are in Jy. Marginalized PDFs are shown, the dashed lines
    marking median values and 0.16 and 0.84 percentiles.
    }
    \label{fig:corner}
\end{figure*}

In Fig.~\ref{fig:corner} we show the Bayesian corner plots for the parameters derived from the fit. The figure also shows the marginalized PDF for each parameter together with the median and 0.16 and 0.84 percentiles. 

\subsection{Ancillary data}

Table~\ref{tab:other} lists the quantities used for producing Fig.~\ref{fig:emy1} and Fig.~\ref{fig:sfr}. Beside the upper limits for $S_{24.6~GHz}^\mathrm{AME}/S_\mathrm{3 THz}$ for our galaxy sample, and  the single-T MBB fits to the reference targets done in this work,  all other data are culled from the literature.

\begin{table*}
\caption{Data of Fig.~\ref{fig:emy1} and Fig.~\ref{fig:sfr}.}
    \label{tab:other}
    \centering
    \small
    \begin{tabular}{lccccccc}
   \hline \hline
   names & distance\tablefootmark{a} & $S_\nu^\mathrm{AME}/S_\mathrm{3 THz}$ & $S_\mathrm{1.4 GHz}$& $T_\mathrm{d}$ & $M_\mathrm{d}$& $M_\star$ & $SFR$ \\
         & Mpc      &                                   & Jy                    & K                 &
   $\mathrm{M_\odot}$ & $\mathrm{M_\odot}$ &
        $\mathrm{M_\odot \,yr^{-1}}$\\ 
         \hline
NGC~3627 & 11.5 & $<2.3\times10^{-4}$~~\tablefootmark{c} & $0.50\pm0.01$\tablefootmark{h}&
25.3$\pm$0.8\tablefootmark{m} & $2.9\pm0.3 \times 10^7$ & $6.7\pm1.2 \times 10^{10}$ &
$2.9\pm1.0$  \\
NGC~4254 & 12.9 & $<3.1\times10^{-4}$~~\tablefootmark{c} & 
$0.51\pm0.02$\tablefootmark{h}&
24.3$\pm$0.8\tablefootmark{m} & $2.6\pm0.2 \times 10^7$& $1.3\pm0.2 \times 10^{10}$ &
$5.1\pm0.5$  \\
NGC~4736 & 4.4 & $<1.8\times10^{-4}$~~\tablefootmark{c} & 
$0.295\pm0.005$\tablefootmark{h}& 
26$\pm$1\tablefootmark{m} & $2.7\pm0.4 \times 10^6$ & $2.5\pm0.4 \times 10^{10}$&
$0.55\pm0.19$ \\
NGC~5055 & 9.0 & $<2.2\times10^{-4}$~~\tablefootmark{c} & 
$0.460\pm0.005$\tablefootmark{h}&
21.3$\pm$0.7\tablefootmark{m} & $4.7\pm0.5 \times 10^7$ & $5.9\pm1.1 \times 10^{10}$&
$2.5\pm0.7$ \\ \hline
MW   & - & $3.4\pm0.3\times10^{-4}$~~\tablefootmark{d}& - & $18.3\pm0.3$\tablefootmark{n} & - & - & - \\
LMC  & 0.05\tablefootmark{b} & $1.4\pm0.2\times10^{-4}$~~\tablefootmark{d} & 
$530\pm30$\tablefootmark{i}& $21.1\pm0.4$\tablefootmark{o} & 
$1.5\pm0.1 \times 10^6$& $2.0\pm0.5\times10^9$~~\tablefootmark{q}& $0.4\pm0.2$\\
SMC  & 0.062\tablefootmark{b} & $7.5\pm2.4\times10^{-5}$~~\tablefootmark{d} & 
$42\pm6$\tablefootmark{i}& $20.2\pm0.8$\tablefootmark{o} & 
$2.6\pm0.4 \times 10^5$& $3.2\pm0.8\times10^9$~~\tablefootmark{q}& $0.015\pm0.010$\\
M~31 & 0.79\tablefootmark{b} & $4.2\pm0.6\times10^{-4}$~~\tablefootmark{e}& 
$5.2\pm0.4$\tablefootmark{j}& $17.3\pm0.1$\tablefootmark{o} & 
$3.1\pm0.1 \times 10^7$& $5.5\pm0.1\times10^{10}$~~\tablefootmark{r} & 
$0.2\pm0.1$\tablefootmark{s}\\
M~33 & 0.04\tablefootmark{b} & $<3.5\times10^{-5}$~~\tablefootmark{f}& 
$2.7\pm0.6$\tablefootmark{k} & $18.9\pm0.3$\tablefootmark{o} & 
$5.6\pm0.4 \times 10^6$& $4.5\pm1.5\times10^9$~~\tablefootmark{t} & 
$0.25\pm0.10$\tablefootmark{u}\\
NGC~253 & 3.7 & $<3.1\times10^{-5}$~~\tablefootmark{g} & 
$6.2\pm0.1$\tablefootmark{l} &  $26.2\pm0.2$\tablefootmark{p} & 
$3.3\pm0.1 \times 10^7$ & $7.6\pm0.8\times10^{10}$~~\tablefootmark{v} & 
$5.8\pm0.8$ \\
NGC~3034& 3.6 & $<4.0\times10^{-5}$~~\tablefootmark{g} & 
$8.4\pm0.3$\tablefootmark{l} & $35\pm1$\tablefootmark{p} & 
$6.6\pm0.5 \times 10^6$ & $1.3\pm0.3\times10^{10}$~~\tablefootmark{w} & 
$4.5\pm0.6$\\
NGC~4945& 3.5 & $<4.2\times10^{-5}$~~\tablefootmark{g} & 
$6.4\pm0.2$\tablefootmark{l}
& $26.4\pm0.4$\tablefootmark{m} & $2.1\pm0.3 \times 10^7$ &
$1.4\pm0.6 \times 10^{10}$&
$6.5\pm3.6$\\
\hline
\end{tabular}
\tablefoot{
\tablefoottext{a}{Unless otherwise specified, distances are from the DustPedia database, available at {\tt http://dustpedia.astro.noa.gr/} .}
\tablefoottext{b}{From \citet{ClarkApJ2021}.}
\tablefoottext{c}{From this work and the {\em Herschel} $S_\mathrm{3 THz}$ in DustPedia.}
\tablefoottext{d}{Estimated at 22.8~GHz in PLA16.}
\tablefoottext{e}{Estimated at 30~GHz in \citet{BattistelliApJL2019}.}
\tablefoottext{f}{From the upper limit of \citet{TibbsMNRAS2018} at 30~GHz and $S_\mathrm{3 THz}$ in \citet{ClarkApJ2021}.}
\tablefoottext{g}{From the upper limit of P11 between 23 and $\approx100$~GHz and {\em Herschel} $S_\mathrm{3 THz}$ in DustPedia, with the exception of NGC~3034, for which $S_\mathrm{3 THz}$ was estimated from the SED fit.}
\tablefoottext{h}{T17; for NGC~3627 we used $S_\mathrm{1.36~GHz}$.}
\tablefoottext{i}{\citet{ForMNRAS2018}.}
\tablefoottext{j}{\citet{BattistelliApJL2019}.}
\tablefoottext{k}{\citet{TibbsMNRAS2018}.}
\tablefoottext{l}{P11.}
\tablefoottext{m}{$T_\mathrm{d}$ and $M_\mathrm{d}$ from FIR-submm MBB fits, $M_\star$ and $SFR$ from optical-to-submm fits \citep{NersesianA&A2019}.}
\tablefoottext{n}{This work; MBB fits of the MW diffuse dust emissivity
for $100\le \lambda/~\mu\mathrm{m}\le 850$ from \citet{HensleyApJ2021}.}
\tablefoottext{o}{This work; $T_\mathrm{d}$ and $M_\mathrm{d}$ from MBB fits of {\em Herschel} flux densities from \citet{ClarkApJ2021}.}
\tablefoottext{p}{This work; $T_\mathrm{d}$ and $M_\mathrm{d}$ from MBB fits of {\em Herschel} flux densities in DustPedia.}
\tablefoottext{q}{$M_\star$ and $SFR$ from \citet{SkibbaApJ2012}.}
\tablefoottext{r}{\citet{ViaeneA&A2014}.}
\tablefoottext{s}{\citet{ViaeneA&A2017}.}
\tablefoottext{t}{\citet{CorbelliMNRAS2003}.}
\tablefoottext{u}{\citet{WilliamsMNRAS2019}.}
\tablefoottext{v}{$M_\star$ and $SFR$ from \citet{DeLosReyesApJ2019}.}
\tablefoottext{w}{$M_\star$ and $SFR$ from \citet{NersesianA&A2019}.}
}

\end{table*}

\end{appendix}

\end{document}